\documentclass[a4paper,11pt]{article}

\usepackage[utf8]{inputenc}
\usepackage[numbers,sort&compress]{natbib}

\usepackage[margin=1.0in]{geometry}

\usepackage{graphics}
\usepackage{graphicx}
\usepackage{url}
\usepackage{caption}
\usepackage{amsmath}
\usepackage{hyperref}
\usepackage{placeins}
\usepackage{morefloats}
\usepackage{slashed}
\usepackage[colorinlistoftodos, shadow]{todonotes}
\usepackage{titlesec}
\usepackage{gensymb}
\usepackage{textcomp}
\usepackage{amssymb}
\usepackage{verbatim}

\def\refeq#1{\mbox{(\ref{#1})}}
\def\reffi#1{\mbox{Fig.~\ref{#1}}}

\def\refta#1{\mbox{Table~\ref{#1}}}

\def\refse#1{\mbox{Section~\ref{#1}}}

\def\citere#1{\mbox{Ref.~\cite{#1}}}
\def\citeres#1{\mbox{Refs.~\cite{#1}}}

\newcommand{\GeV}{\unskip\,\mathrm{GeV}}

\newcommand{\TeV}{\unskip\,\mathrm{TeV}}

\def\mathswitch#1{\relax\ifmmode#1\else$#1$\fi}
\def\mathswitchr#1{\relax\ifmmode{\mathrm{#1}}\else$\mathrm{#1}$\fi}
\def\mathswitchit#1{\relax\ifmmode{#1}\else$#1$\fi}

\newcommand{\Pu}{\mathswitchr u}
\newcommand{\Pd}{\mathswitchr d}
\newcommand{\Ps}{\mathswitchr s}
\newcommand{\Pc}{\mathswitchr c}
\newcommand{\Pt}{\mathswitchr t}
\newcommand{\Pp}{\mathswitchr p}

\newcommand{\Pe}{\mathswitchr e}

\newcommand{\PW}{\mathswitchr W}
\newcommand{\PZ}{\mathswitchr Z}

\newcommand{\PH}{\mathswitchr H}

\def\ie{i.e.\ }

\def\cf{c.f.\ }

\newcommand{\rT}{{\mathrm{T}}}

\newcommand{\Pdbar}{{\bar{\Pd}}}
\newcommand{\Pubar}{{\bar{\Pu}}}
\newcommand{\Psbar}{{\bar{\Ps}}}
\newcommand{\Pcbar}{{\bar{\Pc}}}

\newcommand{\recola}{{\sc Recola}}
\newcommand{\collier}{{\sc Collier}}

\newcommand{\Pb}{\mathswitchr b}

\newcommand{\mr}{\mathrm}

\def\bfi{\begin{figure}}
\def\efi{\end{figure}}

\def\draftdate{\relax}
\def\mda{\relax}
\def\mua{\relax}
\def\mla{\relax}
\def\draft{
\def\thtystars{******************************}
\def\sixtystars{\thtystars\thtystars}
\typeout{}
\typeout{\sixtystars**}
\typeout{* Draft mode!
         For final version remove \protect\draft\space in source file *}
\typeout{\sixtystars**}
\typeout{}
\def\draftdate{\today}
\def\mua{\marginpar[\boldmath\hfil$\uparrow$]%
                   {\boldmath$\uparrow$\hfil}%
                    \typeout{marginpar: $\uparrow$}\ignorespaces}
\def\mda{\marginpar[\boldmath\hfil$\downarrow$]%
                   {\boldmath$\downarrow$\hfil}%
                    \typeout{marginpar: $\downarrow$}\ignorespaces}
\def\mla{\marginpar[\boldmath\hfil$\rightarrow$]%
                   {\boldmath$\leftarrow $\hfil}%
                    \typeout{marginpar: $\leftrightarrow$}\ignorespaces}
\def\Mua{\marginpar[\boldmath\hfil$\Uparrow$]%
                   {\boldmath$\Uparrow$\hfil}%
                    \typeout{marginpar: $\uparrow$}\ignorespaces}
\def\Mda{\marginpar[\boldmath\hfil$\Downarrow$]%
                   {\boldmath$\Downarrow$\hfil}%
                    \typeout{marginpar: $\downarrow$}\ignorespaces}
\def\Mla{\marginpar[\boldmath\hfil$\Rightarrow$]%
                   {\boldmath$\Leftarrow $\hfil}%
                    \typeout{marginpar: $\leftrightarrow$}\ignorespaces}
\overfullrule 5pt
\oddsidemargin -15mm
\marginparwidth 29mm
}

\numberwithin{equation}{section}

\begin{document}

\thispagestyle{empty}
\def\thefootnote{\fnsymbol{footnote}}
\setcounter{footnote}{1}
\null
\draftdate\hfill ICCUB-17-016

\vfill
\begin{center}
  {\Large {\boldmath\bf {Next-to-leading-order electroweak corrections
        to the production of three charged leptons plus missing energy at the LHC}
\par} \vskip 2.5em
{\large
{\sc Benedikt Biedermann$^{1}$, Ansgar Denner$^{1}$, Lars Hofer$^{2}$
}\\[2ex]
{\normalsize \it 
$^1$Julius-Maximilians-Universit\"at W\"urzburg, 
Institut f\"ur Theoretische Physik und Astrophysik, \\
Emil-Hilb-Weg 22, 97074 W\"urzburg, Germany
}\\[2ex]
{\normalsize \it 
$^2$Universitat de Barcelona (UB),
Departament de F\'{\i}sica Qu\`antica i Astrof\'{\i}sica (FQA),\\
Institut de Ci\`encies del Cosmo (ICCUB),
08028 Barcelona, Spain}
}
}
\par \vskip 1em
\end{center}\par
\vskip .0cm \vfill {\bf Abstract:} 
\par 
The production of a neutral and a charged vector boson with subsequent
decays into three charged leptons and a neutrino is a very important
process for precision tests of the Standard Model of elementary
particles and in searches for anomalous triple-gauge-boson couplings.
In this article, the first computation of next-to-leading-order
electroweak corrections to the production of the four-lepton final
states $\mu^+\mu^- \Pe^+ \nu_\Pe$, $\mu^+\mu^- \Pe^- \bar\nu_\Pe$,
$\mu^+\mu^- \mu^+\nu_\mu$, and $\mu^+\mu^- \mu^- \bar\nu_\mu$ at the
Large Hadron Collider is presented. We use the complete matrix elements at
leading and next-to-leading order, including all off-shell
effects of intermediate massive vector bosons and virtual photons.
The relative electroweak corrections to the fiducial cross sections
from quark-induced partonic processes vary between $-3\%$ and $-6\%$,
depending significantly on the event selection. At the level of
differential distributions, we observe large negative corrections of
up to $-30\%$ in the high-energy tails of distributions
originating from electroweak Sudakov logarithms.  Photon-induced
contributions at next-to-leading order raise the leading-order
fiducial cross section by $+2\%$.  Interference effects
in final states with equal-flavour leptons are at the permille level for
the fiducial cross section, but can lead to sizeable effects in
off-shell sensitive phase-space regions.

\par
\vskip 1cm
\noindent
\par
\null \setcounter{page}{0} \clearpage
\def\thefootnote{\arabic{footnote}} \setcounter{footnote}{0}

\tableofcontents

\section{Introduction}
\label{se:intro}

Vector-boson pair production belongs to the most important process
classes at the Large Hadron Collider (LHC). Owing to its sensitivity
to the triple-gauge-boson couplings (TGC), it allows for fundamental
precision tests of the Standard Model (SM) of elementary particles.
In particular, WZ production is considered as one of the key processes
in searches for new physics via anomalous TGC.
Moreover, WZ production is an important SM background to many direct
searches for new physics because the corresponding final state with
three charged leptons plus missing energy leads to a relatively clean
signature.

Both the ATLAS and CMS collaboration have measured WZ production at
$7$, $8$ and $13\TeV$ centre-of-mass energy
\cite{Aad:2012twa,Aad:2016ett,Aaboud:2016yus,Khachatryan:2016tgp,
  Khachatryan:2016poo}. Since the most recent determinations of
anomalous TGC from ATLAS data of run~II \cite{ATLAS:2016qzn} are
compatible with the SM prediction, possible new-physics effects are
severely constrained and expected to be found by looking for small
deviations in high-energy tails of differential distributions.  It is
thus of prime importance to have precise theoretical predictions for
this process at hand.

Most of the efforts for improving the theoretical accuracy of WZ
production have been dedicated to perturbative higher-order
calculations in the strong coupling $\alpha_s$. The first
next-to-leading order (NLO) QCD computation treating the W and Z~boson
as on-shell external particles dates back more than two decades
\cite{Ohnemus:1991gb}. Systematic improvements followed, including
leptonic decays \cite{Ohnemus:1994ff}, off-shell effects and spin
correlations \cite{Dixon:1999di,Campbell:1999ah}. Fixed-order
calculations for WZ production have been matched to parton-shower
generators at NLO QCD
\cite{Hamilton:2010mb,Melia:2011tj,Nason:2013ydw}.  Recently, the
first calculation of next-to-next-to-leading order (NNLO) QCD
corrections for the integrated cross section has been completed
\cite{Grazzini:2016swo} and extended to the level of differential
distributions \cite{Grazzini:2017ckn}.

At this level of accuracy, NLO electroweak (EW) corrections, which are
proportional to the electromagnetic coupling $\alpha$, become relevant
as well. On the one hand, naive power counting $O(\alpha)\approx
O(\alpha_s^2)$ suggests that they are of a similar order of magnitude
as the NNLO QCD corrections. On the other hand, EW corrections can be
enhanced by logarithms of EW origin
\cite{Beenakker:1993tt,Beccaria:1998qe,Ciafaloni:1998xg,Kuhn:1999de,Fadin:1999bq,Denner:2000jv}
and may distort differential distributions at large transverse momenta
by several tens of percent.  The latter property is of particular
importance, since these phase-space regions are most sensitive to
effects of new physics.  NLO EW corrections to WZ~production have
first been studied in a logarithmic approximation
\cite{Accomando:2004de}. A full NLO EW computation for on-shell W and
Z~bosons has been presented later on
\cite{Bierweiler:2013dja,Baglio:2013toa} including also photon-induced
corrections. The NLO EW corrections for the complete
four-lepton-production processes, \ie including leptonic vector-boson
decays and irreducible background diagrams, exist so far only for WW
and ZZ production
\cite{Biedermann:2016yvs,Biedermann:2016lvg,Biedermann:2016guo,Kallweit:2017khh},
while corresponding results for WZ production are still missing in the
literature.

The aim of the present article is to fill this gap and to provide
results for the NLO EW corrections to the production of three charged
leptons plus missing energy at the LHC. We consider the four different
and experimentally well-defined final states $ \mu^+\mu^- \Pe^+
\nu_\Pe$, $\mu^+\mu^- \Pe^- \bar\nu_\Pe$, $\mu^+\mu^- \mu^+\nu_\mu$,
and $\mu^+\mu^- \mu^- \bar\nu_\mu$. We use the complete matrix
elements including besides diagrams with intermediate W and Z~bosons
also those with virtual photons as well as background diagrams with
only one possibly resonant vector boson. In addition, we include also
photon-induced contributions at NLO.  Using the complex-mass scheme
\cite{Denner:1999gp,Denner:2005fg,Denner:2006ic} for a consistent
treatment of resonant propagators, our calculation provides NLO EW
predictions for the entire fiducial volume.  We apply acceptance cuts
inspired by those of the experimental collaborations and study the
impact of the corrections on differential observables that are
relevant in TGC searches.

This article is organized as follows: In \refse{se:methods}, some
details of the computation are outlined.  The numerical setup and the
phenomenological results are presented in \refse{sec:results}.
Finally, conclusions are given in \refse{se:conclusion}.

\section{Details of the calculation}\label{se:methods}

\begin{figure}
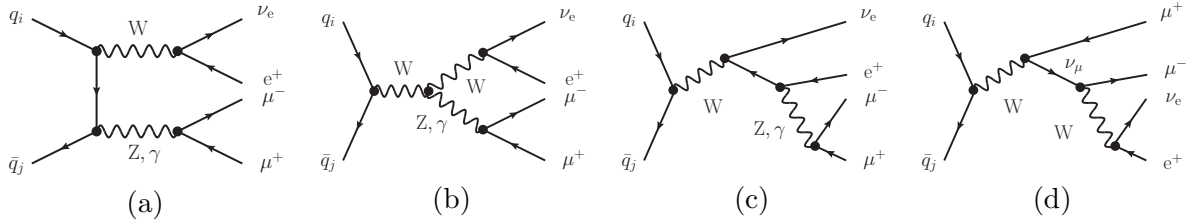

\begin{center}
    \begin{minipage}{0.25\textwidth}
      \includegraphics[width=\textwidth]{diagrams/{{born_QQmmev}}}
      \begin{center} \vspace{-5mm} (a)~~~ \end{center}
    \end{minipage}
    \begin{minipage}{0.24\textwidth}
      \includegraphics[width=\textwidth]{diagrams/{{born_QQmmevTGC}}}
      \begin{center} \vspace{-5mm} (b)~~~ \end{center}
    \end{minipage}
    \begin{minipage}{0.24\textwidth}
      \includegraphics[width=\textwidth]{diagrams/{{born_QQmmevBackground}}}
      \begin{center} \vspace{-5mm} (c)~~~ \end{center}
    \end{minipage}
    \begin{minipage}{0.24\textwidth}
      \includegraphics[width=\textwidth]{diagrams/{{born_QQmmevBackground2}}}
      \begin{center} \vspace{-5mm} (d)~~~ \end{center}
    \end{minipage}
\end{center}
\caption{Sample tree-level diagrams contributing at 
$O(\alpha^4)$ to $q_i\bar{q}_j\to \mu^+\mu^- \Pe^+ \nu_\Pe$.}\label{fig:born}
\end{figure}

We consider the four independent 
processes $\Pp\Pp\to \mu^+\mu^- 
\Pe^+ \nu_\Pe+X$, $\Pp\Pp\to \mu^+\mu^- \Pe^- \bar\nu_\Pe+X$, $\Pp\Pp\to \mu^+\mu^- 
\mu^+\nu_\mu+X$, and $\Pp\Pp\to \mu^+\mu^- \mu^- \bar\nu_\mu+X$. At leading-order 
(LO), the corresponding hadronic cross sections at $O(\alpha^4)$ in the EW 
coupling receive contributions from the following partonic channels:
\begin{align}
 q_i \bar q_j/ \bar q_j q_i&\to \mu^+\mu^- \Pe^+ \nu_\Pe, &  q_i\bar 
q_j\in\{\Pu\Pdbar, \Pc\Psbar, \Pu\Psbar,\Pc\Pdbar\},\nonumber\\
 q_i \bar q_j/ \bar q_j q_i&\to \mu^+\mu^- \mu^+ \nu_\mu, &  q_i\bar 
q_j\in\{\Pu\Pdbar, \Pc\Psbar, \Pu\Psbar,\Pc\Pdbar\},\nonumber\\
 q_i \bar q_j/ \bar q_j q_i&\to \mu^+\mu^- \Pe^- \bar\nu_\Pe, & q_i\bar 
q_j\in\{\Pd\Pubar, \Ps\Pcbar,\Ps\bar\Pu, \Pd\Pcbar\},\nonumber\\
 q_i \bar q_j/ \bar q_j q_i&\to \mu^+\mu^- \mu^-\bar\nu_\mu, &  q_i\bar 
q_j\in\{\Pd\Pubar, \Ps\Pcbar,\Ps\bar\Pu, \Pd\Pcbar\}.\label{eq:qqLO}
\end{align}
We include quark-flavour mixing between the first two quark families
as described by the Cabibbo--Kobayashi--Maskawa (CKM) matrix defined
in Eq.~(\ref{eq:CKMmatrix}), \ie we take into account first-order
mixing but neglect any higher-order quark-flavour mixings.  The
dominant channels involving only quarks and antiquarks of the first
generation contribute about $80\%$ to the integrated LO cross section,
the corresponding channels of the second family between $10\%$ and
$20\%$.
Channels involving quarks of the first and anti-quarks of the second
generation or vice versa, stay at the percent level. Sample tree-level
diagrams contributing to the process $ q_i \bar q_j\to \mu^+\mu^-
\Pe^+ \nu_\Pe$ are shown in \reffi{fig:born}.
Besides diagrams with a resonant W~boson and a resonant Z~boson or
a photon these involve also diagrams with only 
one possibly resonant vector boson.

The NLO EW corrections at $O(\alpha^5)$ comprise virtual corrections
to the partonic channels \refeq{eq:qqLO} as well as real photon emission
via the quark-induced channels
\begin{align}
 q_i \bar q_j/ \bar q_j q_i&\to \mu^+\mu^- \Pe^+ \nu_\Pe\,(+\gamma), & q_i\bar 
q_j\in\{\Pu\Pdbar, \Pc\Psbar, \Pu\Psbar,\Pc\Pdbar\},\nonumber\\
 q_i \bar q_j/ \bar q_j q_i&\to \mu^+\mu^- \mu^+ \nu_\mu\,(+\gamma),& q_i\bar 
q_j\in\{\Pu\Pdbar, \Pc\Psbar, \Pu\Psbar,\Pc\Pdbar\},\nonumber\\
 q_i \bar q_j/ \bar q_j q_i&\to \mu^+\mu^- \Pe^- \bar\nu_\Pe\,(+\gamma),& q_i\bar 
q_j\in\{\Pd\Pubar, \Ps\Pcbar,\Ps\Pubar, \Pd\Pcbar\},\nonumber\\
 q_i \bar q_j/ \bar q_j q_i&\to \mu^+\mu^- \mu^-\bar\nu_\mu\,(+\gamma),& q_i\bar 
q_j\in\{\Pd\Pubar, \Ps\Pcbar,\Ps\Pubar, \Pd\Pcbar\}.\label{eq:qqNLO}
\end{align}
Moreover, we include the photon-induced contributions with one
(anti)quark and one photon in the initial state,
\begin{align}
 \gamma q_i/q_i\gamma &\to \mu^+\mu^- \Pe^+ \nu_\Pe \,q_j, & q_i q_j\in\{\Pu\Pd, 
\Pc\Ps\},\nonumber\\
 \gamma \bar q_i/\bar q_i\gamma &\to \mu^+\mu^- \Pe^+ \nu_\Pe\,\bar q_j ,& \bar 
q_i \bar q_j\in\{\Pdbar\Pubar, \Psbar\Pcbar\},\nonumber\\
 \gamma q_i/q_i\gamma &\to \mu^+\mu^- \mu^+ \nu_\mu \,q_j, & q_i q_j\in\{\Pu\Pd, 
\Pc\Ps\},\nonumber\\
  \gamma \bar q_i/\bar q_i\gamma &\to \mu^+\mu^- \mu^+ \nu_\mu \,\bar q_j, & \bar 
q_i \bar q_j\in\{\Pdbar\Pubar, \Psbar\Pcbar\},\nonumber\\
 \gamma q_i / q_i\gamma&\to \mu^+\mu^- \Pe^- \bar\nu_\Pe \,q_j, & q_i 
q_j\in\{\Pd\Pu, \Ps\Pc\},\nonumber\\
 \gamma \bar q_i / \bar q_i\gamma&\to \mu^+\mu^- \Pe^- \bar\nu_\Pe
 \,\bar q_j, & 
\bar q_i \bar q_j\in\{\Pubar\Pdbar,\Pcbar\Psbar \},\nonumber\\
  \gamma q_i / q_i\gamma&\to\mu^+\mu^- \mu^-\bar\nu_\mu \,q_j, &  q_i 
q_j\in\{\Pd\Pu, \Ps\Pc\},\nonumber\\
  \gamma \bar q_i / \bar q_i\gamma&\to\mu^+\mu^- \mu^-\bar\nu_\mu \,\bar q_j& 
\bar q_i \bar q_j\in\{\Pubar\Pdbar,\Pcbar\Psbar \},\label{eq:qphoNLO}
\end{align}
generically referred to as $q\gamma$ channels in the following.  
In all considered contributions, the bottom quark can neither appear
as initial-state nor final-state particle since its weak isospin
partner, the top quark, is by construction excluded as external
particle in all considered partonic channels at LO and NLO.

Since all considered processes involve exactly one
quark-flavour-changing vertex at tree level and since we treat all
quarks except the top quark as massless, all tree-level amplitudes can
be constructed by multiplying the amplitudes for a unit CKM matrix
with the corresponding non-vanishing CKM matrix elements. This
treatment is also exact for the real and virtual corrections in our
setup. Since the CKM matrix can be eliminated for massless down-type
quarks by a redefinition of flavour eigenstates, no renormalization of
the CKM matrix is required in this approximation.  Non-trivial CKM
effects only enter because the flavour symmetry is broken when
amplitudes for different quark flavours are weighted with different
parton distribution functions (PDFs). For the photon-induced channels
the CKM matrix drops out exactly owing to its unitarity when the sum
over the flavour of the final-state quark is performed. For that
reason, we can restrict the evaluation of the corrections in
\refeq{eq:qphoNLO} to flavour-diagonal channels with the CKM matrix
set to unity.

The one-loop virtual corrections comprise the full set of Feynman
diagrams to the processes \eqref{eq:qqLO} at order $O(\alpha^5)$. Both
at tree and one-loop level, we employ the complex-mass scheme for a
consistent treatment of massive resonant
particles~\cite{Denner:1999gp,Denner:2005fg,Denner:2006ic} leading to
NLO EW accuracy everywhere in phase space. The integration of the real
corrections is performed with help of the subtraction methods of
\citeres{Dittmaier:1999mb,Dittmaier:2008md} in order to deal with soft
and collinear photon emission off fermions and with the collinear
initial-state singularities in the photon-induced corrections. The
employed formalism extends the dipole subtraction from QCD
\cite{Catani:1996vz} to the case with QED splittings. The general idea
of subtraction methods is to add and subtract auxiliary terms that
mimic the singularity structure of the real squared matrix elements
point-wise such that the resulting differences can be integrated in
four space-time dimensions. The re-added subtraction terms, on the
other hand, can be integrated in a process-independent way allowing
for an isolation of the divergences of the real corrections in
analytical form.  For infrared-safe observables, the extracted
collinear final-state singularities and the soft singularities cancel
with the corresponding divergences from the one-loop corrections
according to the Kinoshita--Lee--Nauenberg theorem. The left-over
collinear initial-state singularities are absorbed in redefined
parton-distribution functions. A more detailed description of real and
virtual NLO EW corrections to vector-boson pair production is given in
\citere{Biedermann:2016lvg} for the more general case of the
production of four charged leptons.

The full computation with all possible final states has been performed
with a private Monte Carlo program that has already successfully been
used for the integration of the NLO EW corrections to ZZ and WW
production
\cite{Biedermann:2016yvs,Biedermann:2016lvg,Biedermann:2016guo} and
for the NLO QCD and EW corrections to vector-boson scattering
\cite{Biedermann:2016yds,Biedermann:2017bss}. All the tree-level and
one-loop matrix elements for LO, real and virtual contributions have
been evaluated with the computer program \recola~\cite{Actis:2016mpe}
which internally uses the \collier{} library \cite{Denner:2016kdg} for
the one-loop scalar
\cite{tHooft:1978jhc,Beenakker:1988jr,Dittmaier:2003bc,Denner:2010tr}
and tensor integrals
\cite{Passarino:1978jh,Denner:2002ii,Denner:2005nn}.  As a cross
check, we have performed an independent calculation of the process
$\Pp\Pp\to\mu^+\mu^-\Pe^+\nu_\Pe+X$ and found perfect agreement at the
level of phase-space points and at the level of differential cross
sections within the statistical uncertainty of the Monte Carlo
integration. The matrix elements of the second implementation are
generated with the {\sc Mathematica} package {\sc
  Pole}~\cite{Accomando:2005ra} which is based on {\sc
  FeynArts}~\cite{Kublbeck:1990xc,Hahn:2000kx} and {\sc
  FormCalc}~\cite{Hahn:1998yk}. The phase-space integration is carried
out with an independent multi-channel Monte Carlo integrator based on
the ones described in \citeres{Berends:1994pv,Dittmaier:2002ap}.

\section{Phenomenological results}\label{sec:results}

\subsection{Input parameters}\label{sec:input}

For the numerical analysis we choose the following input parameters
based on \citere{Olive:2016xmw}.  The on-shell masses and widths of
the gauge bosons read
\begin{align}
  \qquad M_\PZ^{\rm os}&=91.1876\GeV,& \Gamma_\PZ^{\rm os}&= 
2.4952\GeV,&\nonumber\\
  \qquad M_\PW^{\rm os}&=80.385\GeV,& \Gamma_\PW^{\rm os}&=2.085\GeV.&
\end{align}
For the use within the complex-mass scheme, they need to be converted to
pole masses and widths 
according to \citere{Bardin:1988xt}:
\begin{align}
 M= \frac{M^{\rm os}}{\sqrt{1+(\Gamma^{\rm os}/M^{\rm os})^2 } }, \qquad\Gamma = 
\frac{\Gamma^{\rm os}}{\sqrt{1+(\Gamma^{\rm os}/M^{\rm os})^2 } }.
\end{align}
For the masses of the Higgs boson and the top quark,  we use
\begin{align}
  M_\PH&=125\GeV,\qquad m_{\Pt}=173\GeV,
\end{align}
while their widths can be set equal to zero as they do not appear as
internal resonances in the considered processes.
Throughout the calculation, all the charged leptons
$\ell=\{\Pe^\pm,\mu^\pm,\tau^\pm\}$ and the five quarks
$q=\{\Pu,\Pd,\Pc,\Ps,\Pb\}$ are considered as light particles with
negligible masses.

The electromagnetic coupling $\alpha$ is derived from the
Fermi constant according to
\begin{align}
 \alpha_{G_\mu}=\frac{\sqrt{2}}{\pi}G_\mu 
 M_\PW^2\left(1-\frac{M_\PW^2}{M_\PZ^2}\right)\qquad {\rm with} \qquad 
G_\mu=1.16637\times 10^{-5}\GeV^{-2},
\end{align}
\ie we work in the $G_\mu$ scheme. In this scheme, the effects of the
running of $\alpha$ from zero-momentum transfer to the electroweak
scale are absorbed into the LO cross section, and mass singularities in
the charge renormalization are avoided.  Moreover, $\alpha_{G_\mu}$
partially accounts for the leading universal renormalization effects
related to the $\rho$-parameter.  We use the following approximation
for the CKM matrix that includes transitions between the first two
quark generations:
\begin{align}
 V_{\rm CKM}=\left(\begin{array}{ccc}
                    V_{\rm u d} & V_{\rm us} & V_{\rm ub} \\
                    V_{\rm c d} & V_{\rm cs} & V_{\rm cb} \\
                    V_{\rm t d} & V_{\rm ts} & V_{\rm tb}
                   \end{array}\right)
            =\left(\begin{array}{ccc}
                    \phantom{-}\cos\theta_{\rm c} & \sin\theta_{\rm c}     & 0 \\
                    -\sin\theta_{\rm c}    & \cos\theta_{\rm c} & 0 \\
                    ~0           & 0           & 1 
                   \end{array}
\right),\qquad\sin\theta_{\rm c}=0.225.\label{eq:CKMmatrix}
\end{align}

Following \citere{Grazzini:2017ckn}, the renormalisation and
factorisation scales, $\mu_{\rm ren}$ and $\mu_{\rm fact}$, are set
equal to the average of the Z-boson and W-boson mass,
\begin{align}
\mu_{\rm ren}=\mu_{\rm fact}=(M_\PZ+M_\PW)/2.
\end{align}
As PDFs we choose the {\tt LUXqed\_plus\_PDF4LHC15\_nnlo\_100}
parameterisation \cite{Manohar:2016nzj,Butterworth:2015oua}.
Throughout our calculation, we employ the $\overline{\rm MS}$
factorisation scheme.  We have numerically verified that the
difference between this scheme and the often used
deep-inelastic-scattering scheme is below one permille for the
relative NLO EW corrections and, thus, phenomenologically irrelevant.

\subsection{Observable definition and acceptance cuts}\label{sec:observablesAndCuts}

Photons emitted in the Bremsstrahlung corrections are recombined with
the closest charged lepton if their separation $\Delta R$ in the
rapidity--azimuthal-angle plane fulfils
\begin{equation}
\label{eq:recombi}
 \Delta R_{\ell_i,\gamma} =
 \sqrt{(y_{\ell_i}-y_\gamma)^2+(\Delta\phi_{\ell_i,\gamma})^2} < 0.1, 
\end{equation}
where $y$ denotes the rapidity of the final-state particle and
$\Delta\phi_{\ell_i,\gamma}$ the azimuthal-angle difference between a
charged lepton $\ell_i$ and the photon $\gamma$. Final-state photons
with rapidity $|y_{\gamma}|>5$ are considered as lost in the beam pipe
and excluded from recombination. Final-state jets present in the
photon-induced real corrections are not recombined with any other
particle.

We refer to $(\ell_{\PZ}^+\ell_{\PZ}^-)$ as the lepton pair associated
with the Z-boson decay and to $\ell_\PW^\pm$ as the charged lepton
from the W-boson decay.  For the processes involving both muons and
electrons, the equal-flavour lepton pair is associated with the
Z-boson decay, while the other charged lepton is associated with the
W-boson decay.  For the processes with three equal-flavour leptons in
the final state, the lepton pair emerging from the Z-boson decay is
defined as the one whose invariant mass $M_{\ell_i^+,\ell_j^-}$ is
closer to the nominal Z-boson mass.

We have investigated each process class in two different
scenarios: first 
with a minimal set of selection cuts, in the
following referred to as ``inclusive setup", and second 
for a setup inspired by the ATLAS measurements \cite{Aad:2016ett,Aaboud:2016yus}
that is tailored to the investigation of 
TGC, referred to as ``TGC setup". The corresponding fiducial volumes
are defined as follows:

\paragraph{Inclusive setup:} We treat all charged final-state leptons
on the same footing, requiring
\begin{align}
p_{\rT,\ell_i}> 15\GeV,\qquad |y_{\ell_i}|&<2.5, \qquad \Delta 
R_{\ell_i,\ell_j} >0.2\label{eq:inc-cuts1},
\end{align}
where $p_{\rT}$ denotes the transverse momentum.

\paragraph{Exclusive setup for TGC analysis:} For each charged lepton
$\ell_i$, we demand a minimal transverse momentum and a maximal
rapidity:
\begin{align}
p_{\rT,\ell_\PZ}> 15\GeV,\qquad p_{\rm 
T,\ell_\PW}>20\GeV,\qquad|y_{\ell_i}|&<2.5. \label{eq:tgc-cuts-pty}
\end{align}
Any pair of charged leptons $(\ell_i,\ell_j)$ is required to be well separated 
in the rapidity--azimuthal-angle plane:
\begin{equation}
\label{eq:tgc-cuts1}
 \Delta R_{\ell_Z,\ell_Z} >0.2,\qquad\Delta R_{\ell_Z,\ell_W} >0.3.
\end{equation}
The invariant mass of the $\ell_{\PZ}^+\ell_{\PZ}^-$~pair is allowed
to differ by at most $10\GeV$ from the nominal Z-boson mass:
\begin{align}\label{eq:invmasscut}
  M_\PZ-10\GeV<\,& M_{\ell_\PZ^+,\ell_\PZ^-}\,<M_\PZ+10\GeV.
\end{align}
The W-boson transverse mass $M_{\rT}^\PW$ must obey
\begin{align}\label{eq:mtwmasscut}
 M_{\rT}^{\PW}=\sqrt{2p_{\rT}^{\rm miss} p_{\rm 
T, \ell_\PW}[1-\cos\Delta\phi(\ell_\PW,\vec{p}_{\rT}^{\rm\, 
miss})]}>30\GeV,
\end{align}
where $\Delta\phi(\ell_\PW,\vec{p}_{\rT}^{\rm\, miss})$ denotes the
azimuthal angle between the momentum of the W-boson decay lepton
$\ell_\PW$ and the missing momentum in the transverse plane
$\vec{p}_{\rT}^{\rm\, miss}$, and $p_{\rT}^{\rm
  miss}=|\vec{p}_{\rT}^{\rm\, miss}|$.

We define the missing momentum as the negative vector sum of the
momenta of all observed particles. Final-state quarks from the
photon-induced corrections are considered as observable jets if their
transverse momentum satisfies
\begin{align}
  p_{\rT}^{\rm jet}>p_{\rT,\rm min}^{\rm jet}=25\GeV.
\end{align}
Hence, quarks in the final state with transverse momentum below
$p_{\rT,\rm min}^{\rm jet}$ contribute to the missing momentum.  All
photons and jets from real radiation with rapidity $|y_{\gamma/\rm
  jet}|>5$ are considered as lost in the beam pipe, and their
four-momentum, thus, contributes to the missing momentum as well. We
note that with this definition, the missing momentum coincides with
the neutrino momentum at LO but not necessarily at NLO.

\subsection{Results on integrated cross sections}\label{sec:results-xsec}



\begin{table}
\begin{center}
\begin{tabular}
{|ll|c|ccc|}
\hline\vspace{-4mm}
&&&&&\\  
\multicolumn{2}{|l|}{mixed flavour $[2\mu\Pe\nu]$}&
$\rule[-1.4ex]{0ex}{2ex} \sigma_\mr{LO}$~[fb] &  $\delta_{\bar q q^\prime}(\%)$&  $\delta_{q \gamma}(\%)$ & $\delta_\mr{NLO}(\%)$
\\
\hline
\hline 
inclusive & $\Pp\Pp \to\mu^+\mu^-\Pe^+\nu_\Pe+X$ & 27.303(1)~ & $-3.308(5)$ & $+1.9564(4)$ & $-1.351(5)$
\\
inclusive & $\Pp\Pp \to\mu^+\mu^-\Pe^-\bar\nu_\Pe+X$ & 17.9133(7) & $-3.211(5)$ & $+2.1004(4)$ & $-1.111(5)$
\\
TGC & $\Pp\Pp \to\mu^+\mu^-\Pe^+\nu_\Pe+X$ & 19.1625(6) & $-5.986(6)$ & $+1.6971(3)$ & $-4.289(6)$
\\
TGC & $\Pp\Pp \to\mu^+\mu^-\Pe^-\bar\nu_\Pe+X$ & 12.8624(4) & $-5.950(6)$ & $+1.7908(3)$ & $-4.160(6)$
\\
\hline
\hline\vspace{-4mm}
&&&&&\\ 
\multicolumn{2}{|l|}{equal flavour $[3\mu\nu]$} & 
\rule[-1.4ex]{0ex}{2ex}$\sigma_\mr{LO}$~[fb] &  $\delta_{\bar q q^\prime}(\%)$&  $\delta_{q \gamma}(\%)$ & $\delta_\mr{NLO}(\%)$
\\
\hline
\hline 
inclusive & $\Pp\Pp \to\mu^+\mu^-\mu^+\nu_\mu+X$ & 27.2448(9) & $-3.310(4)$ & $+1.9577(4)$ & $-1.352(4)$
\\
inclusive & $\Pp\Pp \to\mu^+\mu^-\mu^-\bar\nu_\mu+X$ & 17.8621(6) & $-3.203(4)$ & $+2.1030(4)$  & $-1.100(4)$
\\
TGC  & $\Pp\Pp \to\mu^+\mu^-\mu^+\nu_\mu+X$ & 19.4353(6) & $-5.709(5)$ & $+1.7155(4)$ & $-3.993(5)$
\\
TGC & $\Pp\Pp \to\mu^+\mu^-\mu^-\bar\nu_\mu+X$ & 13.0398(4) & $-5.661(5)$ & $+1.8101(3)$ & $-3.851(5)$
\\
\hline
\end{tabular}
\end{center}
\caption{Fiducial cross sections with first-order quark-flavour
  mixings for all considered final states in the inclusive and TGC setup.} 
\label{tab:all-xsec}
\end{table}

The results for the fiducial cross sections at a centre-of-mass energy
of $13\TeV$ are presented in \refta{tab:all-xsec} for all considered
final states both in the inclusive and in the TGC setup. The second
column shows the absolute prediction for the cross section at leading
order, $\sigma_\mr{LO}$, followed by the relative EW corrections of
the quark-induced contributions $\delta_{\bar q q^\prime}$, the
relative photon-induced corrections $\delta_{q \gamma}$, and the total
relative EW corrections $\delta_\mr{NLO}=\delta_{\bar q
  q^\prime}+\delta_{q \gamma}$. According to the total electric charge
of the final-state leptons, we sometimes refer to the processes
$\Pp\Pp\to \mu^+\mu^- \Pe^+ \nu_\Pe+X$ and $\Pp\Pp\to \mu^+\mu^-
\mu^+\nu_\mu+X$ as $\PZ\PW^+$ and to the processes $\Pp\Pp\to
\mu^+\mu^- \Pe^- \bar\nu_\Pe+X$ and $\Pp\Pp\to \mu^+\mu^- \mu^-
\bar\nu_\mu+X$ as $\PZ\PW^-$. We stress, however, that we include all
contributions leading to the considered four-lepton final state, also
those which do not proceed through intermediate $\PZ\PW^\pm$
production.  The cross sections for the $\PZ\PW^+$ channels are about
$50\%$ larger than the ones for the $\PZ\PW^-$ channels, both in the
mixed-flavour case $[2\mu\Pe\nu]$ and in the equal-flavour case
$[3\mu\nu]$. This can be attributed to the parton flux within the
proton which is larger for the up quark than for the down quark.
The fiducial cross section in the TGC setup is about $30\%$ smaller
than in the inclusive setup, as expected owing to the reduced fiducial
volume. For all setups and channels, the photon-induced contributions
are of the order of $+2\%$ with only minor variations at the
subpercent level. The quark-induced EW corrections are negative and
depend significantly on the phase-space cuts. The corrections in the
TGC setup are with about $-6\%$ almost twice as large as in the
inclusive setup where they reach about $-3\%$. The main reason for the
large difference is the invariant-mass cut in
Eq.~\refeq{eq:invmasscut} which partially removes the radiative tail
below the Z-boson invariant mass as illustrated in the next section in
the context of differential distributions (\cf \reffi{fig:invz}).
Switching off the invariant-mass cut in the TGC setup leads to EW
corrections in the quark-induced channels of $\sim-3.5\%$, \ie much
closer to the results from the inclusive setup. Owing to the opposite
sign of the photon-induced and quark-induced corrections, the net
corrections to the fiducial cross section are only about $-1\%$ in the
inclusive setup and remain around $-4\%$ in the TGC setup.

The inclusion of first-order transitions in flavour-changing currents
lowers the total cross section with respect to a unit CKM matrix by
$0.7\%$ in the $\PZ\PW^+$ channels and by $0.9\%$ in the $\PZ\PW^-$
channels independently of the leptons in the final state.  We have
also performed a LO study including transitions between the second and
third quark generation, which prooved that this effect is
phenomenologically irrelevant.

Since the cuts of the inclusive setup are by construction not
sensitive to the lepton pairing, the scenario is well suited to study
the size of interference effects present for equal-flavour leptons in
the final state. In the absence of any interference, the equal-flavour
and mixed-flavour cross sections would be equal.  The deviation of the
ratio $\sigma^{[3\mu\nu]}/\sigma^{[2\mu\Pe\nu]}$ from one thus gives a
measure of the impact of interferences. At LO, we find
$\sigma^{\mu^+\mu^-\mu^+\nu_\mu}/\sigma^{\mu^+\mu^-\Pe^+\nu_\Pe}=0.99785(5)$
and
$\sigma^{\mu^+\mu^-\mu^-\bar\nu_\mu}/\sigma^{\mu^+\mu^-\Pe^-\bar\nu_\Pe}
=0.99714(5)$. 
Hence, the interferences lower the LO cross sections at the permille
level. The interference effect on the relative NLO EW corrections is
far below the permille level and phenomenologically unimportant for
the fiducial cross section. We conclude that, in an inclusive
scenario, the theory prediction for the integrated NLO EW cross
section can be covered both for the mixed-flavour and equal-flavour
final state by a single computation. In the TGC scenario, the
interference effects cannot be isolated uniquely as the phase-space
cuts are not symmetric under the exchange of the two identical
final-state leptons.  It is thus not surprising to find larger
deviations of $1.4\%$ from the unit ratio
[$\sigma^{\mu^+\mu^-\mu^+\nu_\mu}/\sigma^{\mu^+\mu^-\Pe^+\nu_\Pe}=1.01424(5)$
and
$\sigma^{\mu^+\mu^-\mu^-\bar\nu_\mu}/\sigma^{\mu^+\mu^-\Pe^-\bar\nu_\Pe}
=1.01379(5)$].

We conclude this section with a comparison of our results for the
fiducial cross section with those in the literature for on-shell WZ
production. The computation of \citere{Baglio:2013toa} includes
photon-induced and quark-induced corrections, and states for the total
cross section (no phase-space cuts applied) a negligible EW
correction.  Unfortunately, there are no separate numbers for the
$q\gamma$ and $\bar qq^\prime$ channels for a detailed comparison.
The computation in \citere{Bierweiler:2013dja} does not include
photon-induced corrections. For the LHC at $14\TeV$, the authors state
corrections of $\delta_{\bar q q^\prime}= -1.5\%$ for $\PZ\PW^+$ and
$\delta_{\bar q q^\prime}= -1.3\%$ for $\PZ\PW^-$, applying a minimal
event selection that is roughly comparable with our inclusive setup.
We attribute the difference of $2\%$ to our results
mainly to photon radiation off the $\mu^+\mu^-$ pair.  Inspection of
the (unmeasurable) four-lepton invariant-mass distribution reveals
that above the pair-production threshold, where the cross section
receives the largest contribution, the NLO EW corrections are negative
at the level of $-3\%$ and dominated by real photon radiation. For an
on-shell Z~boson, the effect of final-state radiation and thus of real
corrections is reduced.

\subsection{Results on differential cross 
sections}\label{sec:differentialXsections}

In the following, we present results for distributions for the LHC at
$13\TeV$. In each of the figures, the upper panels show the absolute
predictions for the LO and NLO differential cross section while the
lower panels display the relative EW corrections.

We first discuss the mixed-flavour final state where the $\mu^+\mu^-$
pair can be associated with the decay of the neutral vector boson,
distinguishing between the $\PZ\PW^+$ and $\PZ\PW^-$ case.
Figure~\ref{fig:invz} shows the invariant-mass distributions of the
$\mu^+\mu^-$ system.
\bfi
\begin{center}
\begin{minipage}{0.49\textwidth}
 \includegraphics[width=\textwidth]{plots/{{inc.13TeV.invz.highres.2m1e}}}
\end{minipage}
\begin{minipage}{0.49\textwidth} 
  \includegraphics[width=\textwidth]{plots/{{ATLAStgc.13TeV.invz.highres.2m1e}}}
\end{minipage}
\end{center}
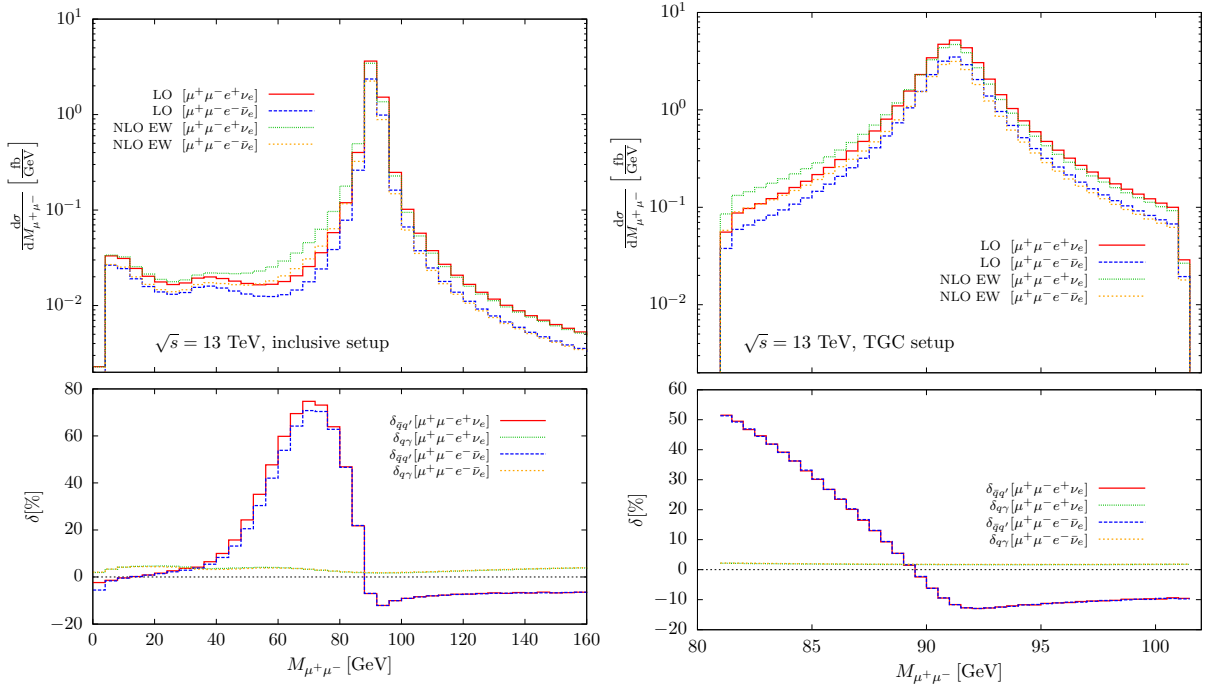
\caption{Distribution in the invariant mass of the $\mu^+\mu^-$ pair
  in the inclusive setup (left panel) and in the TGC setup (right panel).}
\label{fig:invz}
\efi
The absolute prediction in the inclusive setup (left panel) exhibits
the characteristic pattern of this observable similar to the
corresponding $\mu^+\mu^-$ invariant-mass distribution in ZZ
production in \citere{Biedermann:2016lvg}: 1) the resonance peak at
$M_{\mu^+\mu^-}=M_\PZ$, 2) the increase of the cross section towards
$M_{\mu^+\mu^-}=0$ owing to the tail of the photon pole, and 3) a
little bump between $30\GeV$ and $50\GeV$ from the $s$-channel
resonance at $M_{\mu^+\mu^-\Pe^\pm\overset{\smash{(-)}}{\nu}_\Pe}=M_\PW$ [\cf
diagrams (b), (c) and (d) in \reffi{fig:born}].  Turning to the EW
corrections, we observe in the quark-induced channels a typical
radiative tail with corrections of up to $+75\%$: Photon radiation off
the final-state charged leptons may shift the measured value of the
invariant mass to lower values. Since the LO cross section falls off
steeply below the resonance, the relative real NLO corrections become
large. At the resonance the corrections change sign, and above they are
of the order of $-10\%$ (they reach $-25\%$ at $1.5\TeV$, not shown in
the plot). The relative EW corrections are almost equal for $\PZ\PW^+$
and $\PZ\PW^-$, the only visible difference being in the radiative
tail which is up to $6\%$ larger for $\PZ\PW^+$.
This difference results from  folding the partonic cross
sections with the PDFs.
The photon-induced corrections are positive over the whole spectrum
with variations between $1.8\%$ and $5\%$.
Owing to the cut around the Z-boson resonance \refeq{eq:invmasscut},
the invariant $\mu^+\mu^-$ mass in the TGC scenario (right panel) is
restricted to $[M_\PZ-10,M_\PZ+10]$. Evidently, this cut removes a
substantial part of the radiative tail present in the inclusive setup.

Figure \ref{fig:mtwz} shows the distribution in the transverse mass
$M_{\rT}^{3\ell\nu}$ of the four-lepton system in the inclusive
setup (left panel) and in the TGC setup (right panel) as defined in
\citere{Aad:2016ett},
\begin{align}
  M_{\rT}^{3\ell\nu}=\sqrt{\left(\sum_{\ell_i=1}^3 p_{\rm 
T,\ell_i}+|\vec{p}_{\rT}^{\,\rm miss}|\right)^2
      -\left[\left(\sum_{\ell_i=1}^3p_{\ell_i,_x}+p_x^{\rm miss}\right)^2
      +\left(\sum_{\ell_i=1}^3p_{\ell_i,_y}+p_y^{\rm 
miss}\right)^2\,\right]}\label{obs:mtwz}
\end{align}
with $\ell_1=\ell_\PZ^+$, $\ell_2=\ell_\PZ^-$, $\ell_3=\ell_\PW^\pm$
and the missing momentum $\vec{p}^{\,\rm miss}$ defined at the end of
\refse{sec:observablesAndCuts}. 
\bfi
\begin{center}
\begin{minipage}{0.49\textwidth}
 \includegraphics[width=\textwidth]{plots/{{inc.13TeV.mtwz.2m1e}}}
\end{minipage}
\begin{minipage}{0.49\textwidth}
  \includegraphics[width=\textwidth]{plots/{{ATLAStgc.13TeV.mtwz.2m1e}}}
\end{minipage}
\end{center}
\caption{Distribution in the transverse mass of the four-lepton system
  in the inclusive setup (left panel) and in the TGC setup (right
  panel).}
\label{fig:mtwz}
\efi
Note that for contributions with only leptons in the final state, like
the virtual corrections or the LO contribution, the transverse mass in
Eq.~\refeq{obs:mtwz} reduces to the scalar sum of the lepton
transverse momenta. The absolute prediction has its maximum slightly
below $M_{\rT}^{3\ell\nu}=M_\PZ+M_\PW$. The observable does not show a
sharp pair-production threshold (like the unmeasurable four-lepton
invariant-mass distribution would exhibit at $M_{4\ell}=M_\PZ+M_\PW$,
\cf the discussion in \citere{Biedermann:2016lvg} for ZZ production)
as the unmeasurable boost of the four-lepton system along the beam
axis allows for on-shell production of the W and Z~boson below
$M_{\rT}^{3\ell\nu}=M_\PZ+M_\PW$. The little peak directly below
$80\GeV$ in the inclusive setup stems from a single W-boson resonance
with
$M_\PW^2=(p_{\ell_\PZ^+}+p_{\ell_\PZ^-}+p_{\ell_\PW^\pm}+p_\nu)^2$
[\cf diagrams (b), (c) and (d) in \reffi{fig:born}]. This resonance is
removed in the TGC setup owing to the lower cut on the invariant
$\mu^+\mu^-$ mass in Eq.~\refeq{eq:invmasscut} and the minimal
transverse momentum $p_{\rT,\ell_\PW}^{\rm min}$ of the charged
W-decay lepton candidate in Eq.~\refeq{eq:tgc-cuts-pty} since
$(M_\PZ-10\GeV)+p_{\rT,\ell^\PW}^{\rm min}>M_\PW$. The shape of the
quark-induced EW corrections above the maximum of the distribution
is very similar in both setups. We observe a plateau region from the
maximum on up to about $300\GeV$ with $-5\%$ corrections in the
inclusive setup ($-7\%$ in the TGC setup) and then a constant decrease
to $-20\%$ ($-22\%$) at $1\TeV$.  The shape of the NLO EW corrections
can be understood best by analysing the contributions from the
subtracted virtual and real corrections separately. Up to about
$250\GeV$, the subtracted virtual corrections contribute less than one
percent, above $250\GeV$, however, they grow negative with constant
slope and dominate the entire high-energy behaviour owing to EW
Sudakov logarithms. Above the
maximum of the distribution, the subtracted real corrections cause a
flat off-set and only slightly decrease in magnitude with growing
$M_{\rT}^{3\ell\nu}$.  The combination of the virtual and real
corrections gives rise to the plateau in the distribution. The region
below the maximum is entirely dominated by the subtracted real
corrections: the kink at the maximum followed by increasing
corrections is due to the radiative return of the real photon at the
relatively broad peak. The difference between the TGC and the
inclusive setup in this region results from the enhanced radiative
tail from the reconstructed Z-boson resonance, as we checked
explicitly by switching off the invariant-mass cut in
Eq.~\refeq{eq:invmasscut}.  The photon-induced corrections have their
minimum with about $1\%$ where the LO quark-induced channels are
largest, and constantly increase with growing $M_{\rT}^{3\ell\nu}$ up
to $5\%$ to $8\%$, depending on the final state and the setup.

\bfi
\begin{center}
\begin{minipage}{0.49\textwidth}
  \includegraphics[width=\textwidth]{plots/{{inc.13TeV.mtwz.3l}}}
\end{minipage}
\begin{minipage}{0.49\textwidth}
  \includegraphics[width=\textwidth]{plots/{{ATLAStgc.13TeV.mtwz.3l}}}
\end{minipage}
\end{center}
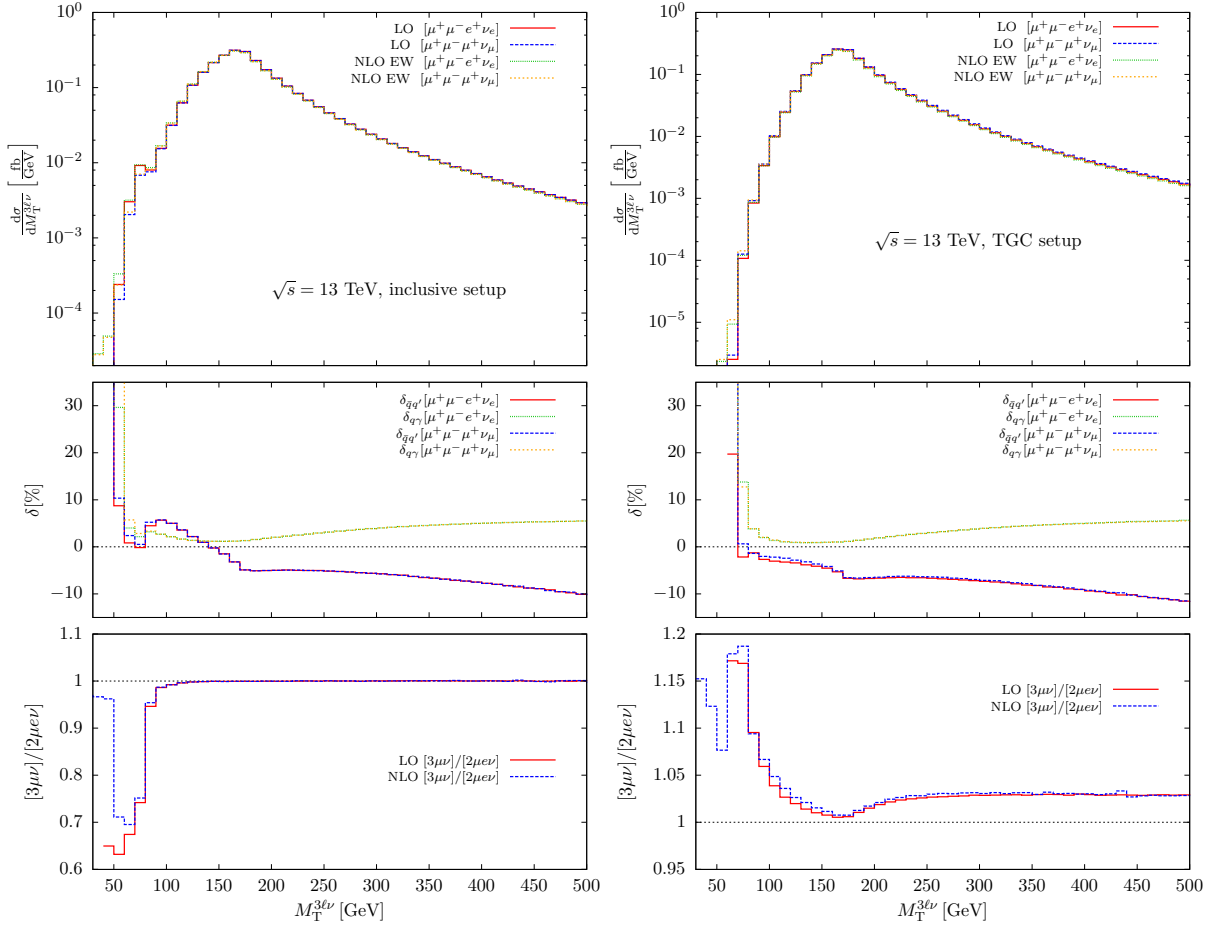
\caption{Comparison between equal-flavour and mixed-flavour 
final state for the distribution in the four-lepton transverse mass 
$M^{3\ell\nu}_{\rT}$ for $\PZ\PW^+$ production. The left 
panel shows the inclusive setup, the right panel the TGC setup.}
\label{fig:mtwz.3l}
\efi
Figure~\ref{fig:mtwz.3l} compares the transverse mass
$M^{3\ell\nu}_{\rT}$ of the four-lepton system for the equal-flavour
$[3\mu\nu]$ and mixed-flavour $[2\mu\Pe\nu]$ final states in
$\PZ\PW^+$ production for the inclusive setup (left panel) and for the
TGC setup (right panel). For both scenarios, the relative EW
corrections of the $\mu^+\mu^-\Pe^+\nu_\Pe$ and
$\mu^+\mu^-\mu^+\nu_\mu$ final states are almost equal (separately for
the $\bar qq^\prime$ and $q\gamma$ channels). This is in agreement
with the results for the fiducial cross section where only
permille-level differences between corrections of the mixed- and
equal-flavour final states are observed. In the lowest panel, the ratio
$({\rm d}\sigma_{\rm (N)LO}^{[3\mu\nu]}/{\rm d}{M_{\rT}^{3\ell
    \nu}})/({\rm d}\sigma_{\rm (N)LO}^{[2\mu\Pe\nu]}/{\rm
  d}{M_{\rT}^{3\ell \nu}})/$ is shown. By construction, the observable
$M^{3\ell\nu}_{\rT}$ is not sensitive to the assignment of the decay
leptons to the Z or the W~boson. Since the inclusive setup is
symmetric in the equal-flavour final-state leptons, the deviation from
one of this ratio gives a direct measure of the impact of
interferences. In the off-shell-sensitive region below $100\GeV$, the
interferences become indeed sizeable, lowering the $[3\mu\nu]$ cross
section by about one third with respect to the $[2\mu\Pe\nu]$ case. As
expected, the interferences are irrelevant in the on-shell region,
where they are suppressed with respect to the doubly-resonant
contributions. In the TGC setup we observe relative differences
between the two final states of the order of $2{-}3\%$ also in the
on-shell region. Because of the smallness of the interference effects
in the inclusive setup we attribute the deviation from one in the
on-shell region in the TGC setup to the lepton pairing in the presence
of asymmetric cuts on $\ell_\PW$ and $\ell_\PZ$.  Around the maximum,
the ratio deviates from one at the percent level, in agreement with
the ratio for the fiducial cross section.  Further below the maximum,
a separation of interference and lepton-pairing effects is not
possible.  In both setups, the NLO EW corrections do not modify the
shape of the ratio.

\bfi
\begin{center}
\begin{minipage}{0.49\textwidth}
  \includegraphics[width=\textwidth]{plots/{{ATLAStgc.13TeV.mtw.2m1e}}}
\end{minipage}
\begin{minipage}{0.49\textwidth} 
  \includegraphics[width=\textwidth]{plots/{{ATLAStgc.13TeV.ptz.2m1e}}}
\end{minipage}
\end{center}
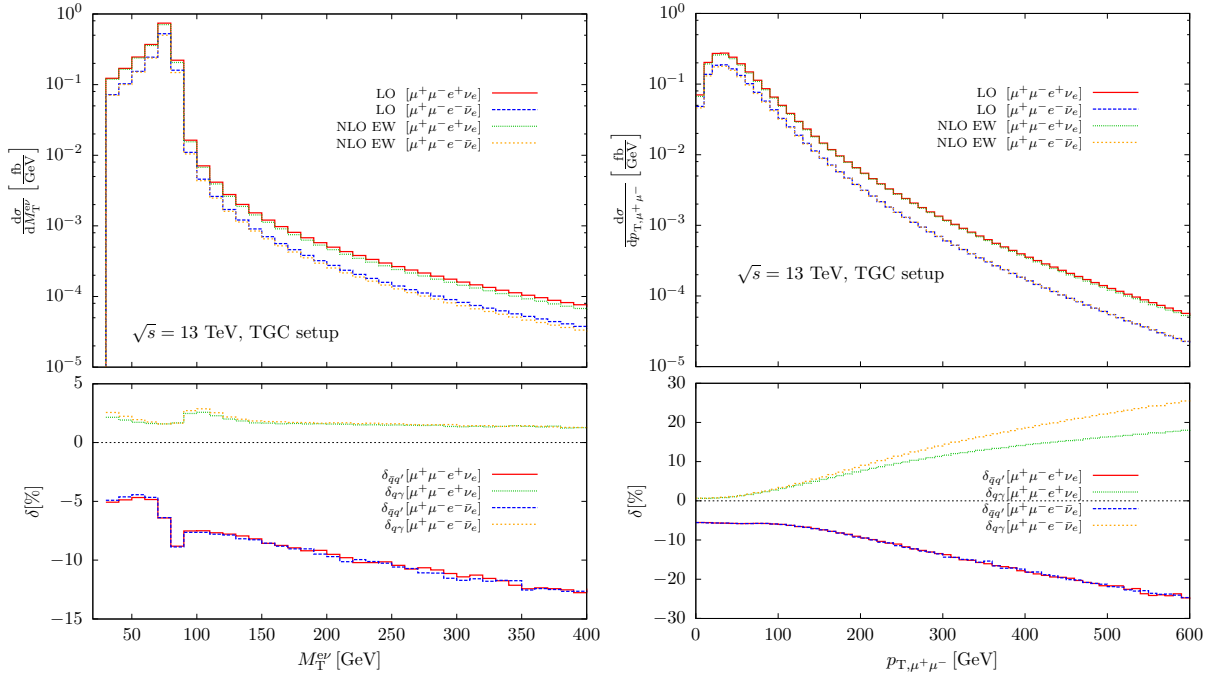
\caption{Distribution in the transverse mass of the reconstructed W~boson (left panel) and  
in the transverse momentum of the $\mu^+\mu^-$ pair (right panel) in
the TGC setup.}
\label{fig:mtw+ptz}
\efi
The left plot in \reffi{fig:mtw+ptz} shows the distribution in the
transverse mass of the reconstructed W~boson,
$M_{\rT}^{\PW}=M_{\rT}^{\Pe \nu}$, as defined in
Eq.~\refeq{eq:mtwmasscut} in the TGC setup.  The peak of the
distribution is located below the W-boson mass (the reconstructed
invariant W-boson mass is experimentally not accessible owing to the
undetected neutrino). The quark-induced corrections follow a similar
pattern as already observed in $M_{\rT}^{3\ell\nu}$: Below the
maximum, the subtracted virtual corrections are small (less than $1\%$
in magnitude). Above, they constantly increase in size up to $-7.5\%$
with growing $M_{\rT}^{\Pe\nu}$ due to logarithms of EW origin. The
subtracted real corrections above $100\GeV$ give again an off-set of
the order of $-5\%$. Both the dip right above the maximum and the
increase below that maximum represent the radiative response of the
broadly peaked transverse-mass distribution: Final-state radiation off
the W-decay lepton shifts the numerical value of the observable closer
to (further away from) the maximum. Depending on the slope of the LO
prediction above (below) the maximum this decreases (increases) the
relative corrections. In the inclusive setup (not shown) the shape of
the distribution is similar to the one in the TGC setup, both for the
absolute prediction and the NLO EW corrections.  The relative
quark-induced corrections in the inclusive setup differ from the ones
in the TGC setup only by a constant shift of about $+2.5\%$.  This is
expected, since the main difference between the setups, the cut around
the $\mu^+\mu^-$ invariant mass in Eq.~\refeq{eq:invmasscut}, does not
directly influence $M_{\rT}^{\Pe\nu}$ as this observable does not
depend on the muon momenta. The photon-induced corrections are
relatively flat and do not show any particularly interesting pattern.

The right plot in \reffi{fig:mtw+ptz} shows the transverse-momentum
distribution of the $\mu^+\mu^-$ system in the TGC setup, \ie the
transverse momentum of the reconstructed Z~boson. We observe the
typical feature of large EW corrections in the quark-induced channels
that reach $-25\%$ at $600\GeV$ due to EW Sudakov logarithms. We can
compare this number with the corresponding results for the
distribution in the Z-boson transverse momentum of the on-shell
calculations of \citeres{Baglio:2013toa,Bierweiler:2013dja}. From the
plots in these references we extract a correction of about $-22\%$ at
$p_{\rT,\PZ}=600\GeV$. We attribute the difference of $-3\%$ to the
slightly different setup and to the missing final-state radiation off
muons (radiative energy loss shifts events to smaller transverse
momentum and thus leads to more negative corrections). Similarly to
the previously considered observables that depend on transverse
momenta, the subtracted virtual corrections are small (below $1\%$) in
the low-$p_{\rT}$ region and start growing negative with constant
slope above $100\GeV$. The corrections in the low-$p_{\rT}$ region,
where the bulk of the cross section stems from, are entirely dominated
by the subtracted real radiation. The fact that the corrections are
flat there is again due to the invariant-mass cut
\refeq{eq:invmasscut}. In the inclusive setup (not shown) the
corrections continuously decrease in size until approaching $-1.5\%$
at
zero transverse momentum. The most remarkable feature of this
observable is the large increase of the photon-induced contributions
for high transverse momenta. At $600\GeV$, they reach $+18\%$ in the
$\PZ\PW^+$ case and even $+25\%$ in the $\PZ\PW^-$ case, and, thus,
almost compensate the large negative EW corrections from the
quark-induced channels. The large difference between $\PZ\PW^+$ and
$\PZ\PW^-$ is caused by the different PDFs involved.  In
\citere{Baglio:2013toa}, it has been shown that the large increase of
the photon-induced cross section is mainly due to the coupling of the
photon to the W~boson. The photon-induced corrections presented in
\citere{Baglio:2013toa} show with $+28\%$ for $\PZ\PW^+$ and $+41\%$
for $\PZ\PW^-$ qualitatively a similar behaviour, though the numerical
values differ. We attribute the difference mainly to the different PDF
set as the ratio between the photon-induced real corrections and the
purely quark-induced LO contribution is very sensitive to the employed
photon PDF. The large photon-induced corrections can be reduced by
imposing a jet veto \cite{Biedermann:2016guo}.

\bfi
\begin{center}
\begin{minipage}{0.49\textwidth}
  \includegraphics[width=\textwidth]{plots/{{ATLAStgc.13TeV.pte.2m1e}}}
\end{minipage}
\begin{minipage}{0.49\textwidth}
  \includegraphics[width=\textwidth]{plots/{{ATLAStgc.13TeV.ptmiss.2m1e}}}
\end{minipage}
\end{center}
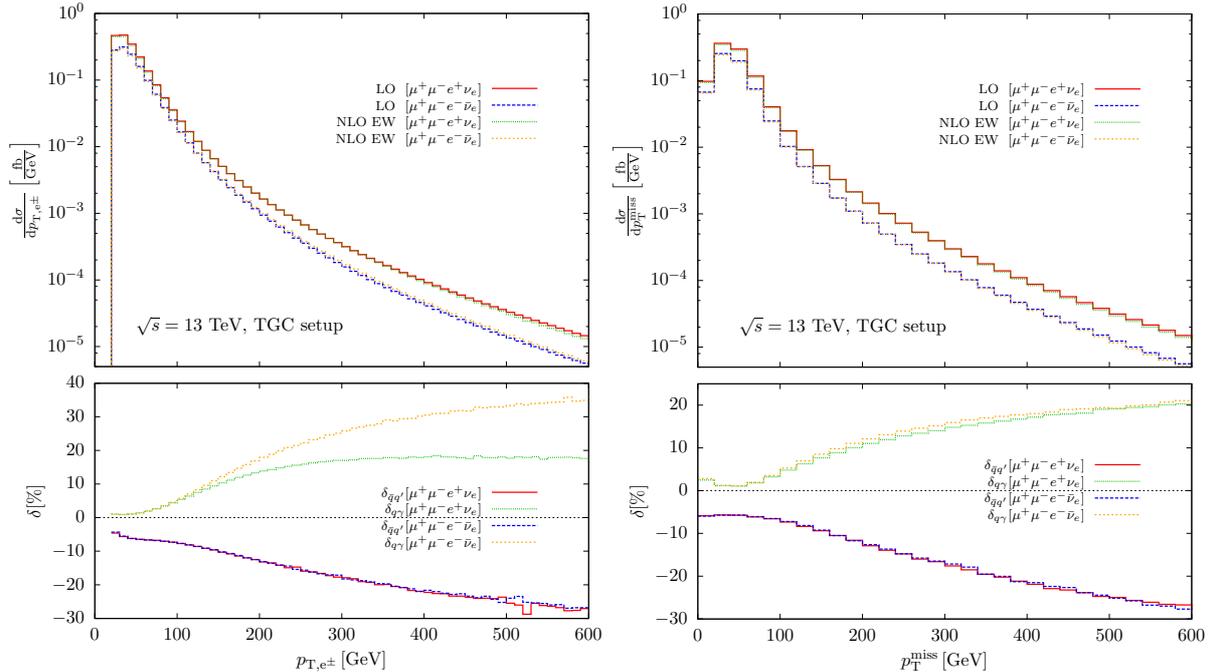
\caption{Distribution in the transverse-momentum of the charged
  W-decay lepton (left panel) and the missing-transverse momentum
  (right panel) in the TGC setup.}
\label{fig:pte.ptmiss}
\efi
The transverse-momentum distributions in \reffi{fig:pte.ptmiss} for
the charged W-decay lepton (left panel) and the missing transverse
momentum (right panel) show similar features as already observed in
the transverse-momentum distribution of the $\mu^+\mu^-$ pair in
\reffi{fig:mtw+ptz}: Large negative EW corrections in the $\bar
qq^\prime$ channel and large positive corrections from the
photon-induced contributions in the high-$p_{\rT}$ regime that
partially compensate each other. Among all transverse-momentum
distributions, those for the transverse momentum of the W-decay lepton
show the largest difference in the photon-induced corrections between
$\PZ\PW^+$ and $\PZ\PW^-$.

The distribution in the azimuthal-angle difference of the $\mu^+\mu^-$
pair in the TGC setup is shown in the left panel of
\reffi{fig:dphi.zleptons} for the $\PZ\PW^+$ and $\PZ\PW^-$
mixed-flavour case. 
\bfi
\begin{center}
\begin{minipage}{0.49\textwidth}
  \includegraphics[width=\textwidth]{plots/{{ATLAStgc.13TeV.dphi.zleptons.2m1e}}}
\end{minipage}
\begin{minipage}{0.49\textwidth}
  \includegraphics[width=\textwidth]{plots/{{ATLAStgc.13TeV.dphi.wzleptons.2m1e}}}
\end{minipage}
\end{center}
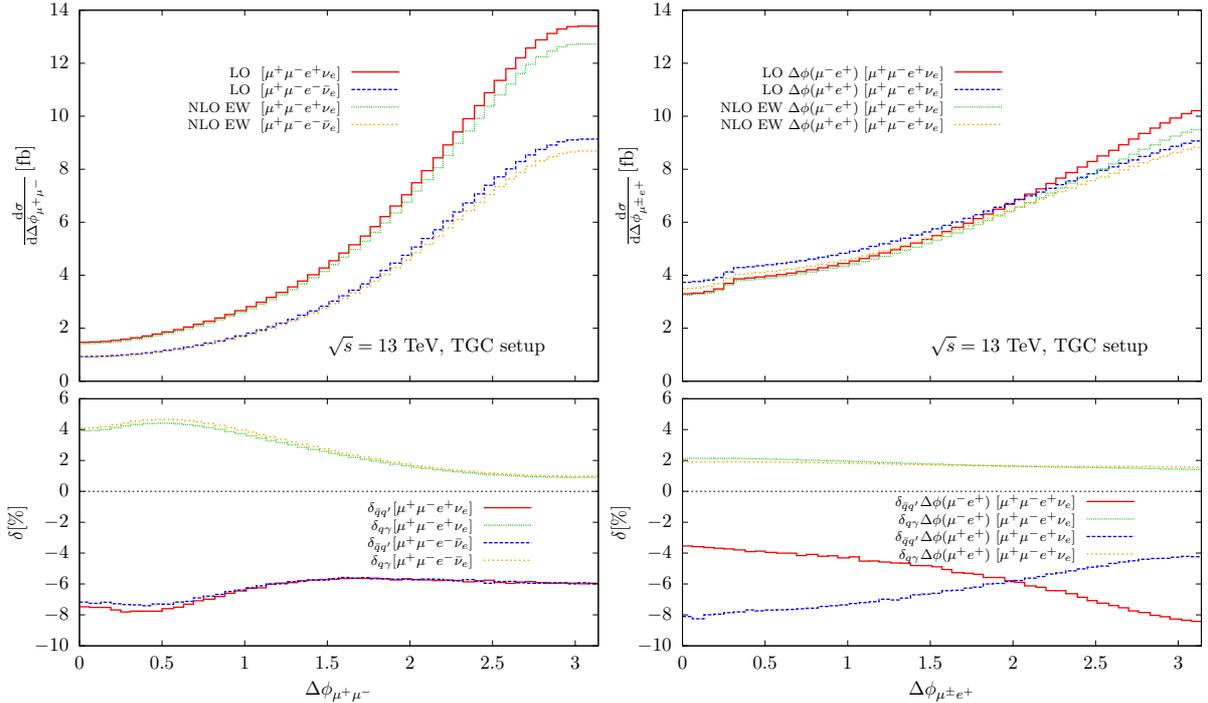
\caption{Distributions in the azimuthal-angle difference of the
  charged leptons in the TGC setup for the mixed-flavour case. The
  left panel shows the correlation between the $\mu^+\mu^-$ pair for
  $\PZ\PW^+$ and $\PZ\PW^-$. In the right panel, the correlations
  between the $\mu^-\Pe^+$ pair and the $\mu^+\Pe^+$ pair are plotted
  for the $\PZ\PW^+$ channel.}
\label{fig:dphi.zleptons}
\efi
The maximum at $\Delta\phi\to\pi$ has the same origin as for the
corresponding observable in ZZ production described in
\citere{Biedermann:2016lvg}: The whole distribution is dominated by
events in the energy region just above the pair-production threshold
with two resonant vector bosons. Owing to the $t$-channel nature of
the dominant contributions [\cf diagram (a) in \reffi{fig:born}], the
vector bosons are preferably produced in forward direction with small
momenta in the transverse plane. The Z-boson decay leptons are thus
mainly back-to-back in the transverse plane which explains the maximum
at $\Delta\phi\to\pi$. The EW corrections are nearly equal for both
presented final states. Around the maximum, the $\bar qq^\prime$
channels receive corrections of $-6\%$ as for the fiducial cross
section.  Towards $\Delta\phi\to0$, the EW corrections are more
enhanced and increase in magnitude up to $-8\%$.  This region is
dominated by events with large transverse momenta of the Z~boson and,
therefore, enhanced by Sudakov logarithms.  The photon-induced
corrections show the opposite behaviour: there is a minimum at
$\Delta\phi=\pi$ of $\sim+1\%$, and a maximum at $\Delta\phi=0.5$ of
$\sim+5\%$.  Hence, both for the $\bar q q^\prime$ and $q\gamma$
channels, the NLO EW corrections to $\Delta\phi_{\mu^+\mu^-}$ reflect
qualitatively the behaviour of the corrections in the
transverse-momentum distribution of the $\mu^+\mu^-$ pair in
\reffi{fig:mtw+ptz}.

The plot on the right-hand side in \reffi{fig:dphi.zleptons} compares
the azimuthal-angle difference of the $\mu^-\Pe^+$ pair with the one
of the $\mu^+\Pe^+$ pair in the mixed-flavour $\PZ\PW^+$ final state.
In both cases we observe a maximum at $\Delta\phi=\pi$, and a minimum
at $\Delta\phi=0$ resulting from boosts of back-to-back W and Z
bosons.  The kink at $\Delta\phi=0.3$ is due to the lepton-separation
cut in Eq.~\refeq{eq:tgc-cuts1}. The difference between maximum and
minimum is much smaller than for $\Delta\phi_{\mu^+\mu^-}$, and
smallest for $\Delta\phi_{\mu^+\Pe^+}$. The photon-induced corrections
show a rather flat behaviour and are practically independent of the
observable. The $\bar q q^\prime$-induced corrections, however, differ
significantly for the two observables: The corrections in the
$\mu^-\Pe^+$ case decrease from $-3.5\%$ at
$\Delta\phi_{\mu^-\Pe^+}=0$ to $-8.4\%$ at
$\Delta\phi_{\mu^-\Pe^+}=\pi$, while those in the $\mu^+\Pe^+$ case
increase from $-8.2\%$ to $-4.2\%$ within the same range of
$\Delta\phi_{\mu^+\Pe^+}$. The observed difference is mainly caused by
the real corrections and due to events close to the WZ production
threshold.

\bfi
\begin{center}
\begin{minipage}{0.49\textwidth}
 \includegraphics[width=\textwidth]{plots/{{ATLAStgc.13TeV.rap.mup.2m1e}}}
\end{minipage}
\begin{minipage}{0.49\textwidth}
  \includegraphics[width=\textwidth]{plots/{{ATLAStgc.13TeV.rap.mum.2m1e}}}
\end{minipage}
\end{center}
\begin{center}
\begin{minipage}{0.49\textwidth}
 \includegraphics[width=\textwidth]{plots/{{ATLAStgc.13TeV.rap.zleptons.2m1e}}}
\end{minipage}
\begin{minipage}{0.49\textwidth}
  \includegraphics[width=\textwidth]{plots/{{ATLAStgc.13TeV.rap.wlepton.2m1e}}}
\end{minipage}
\end{center}
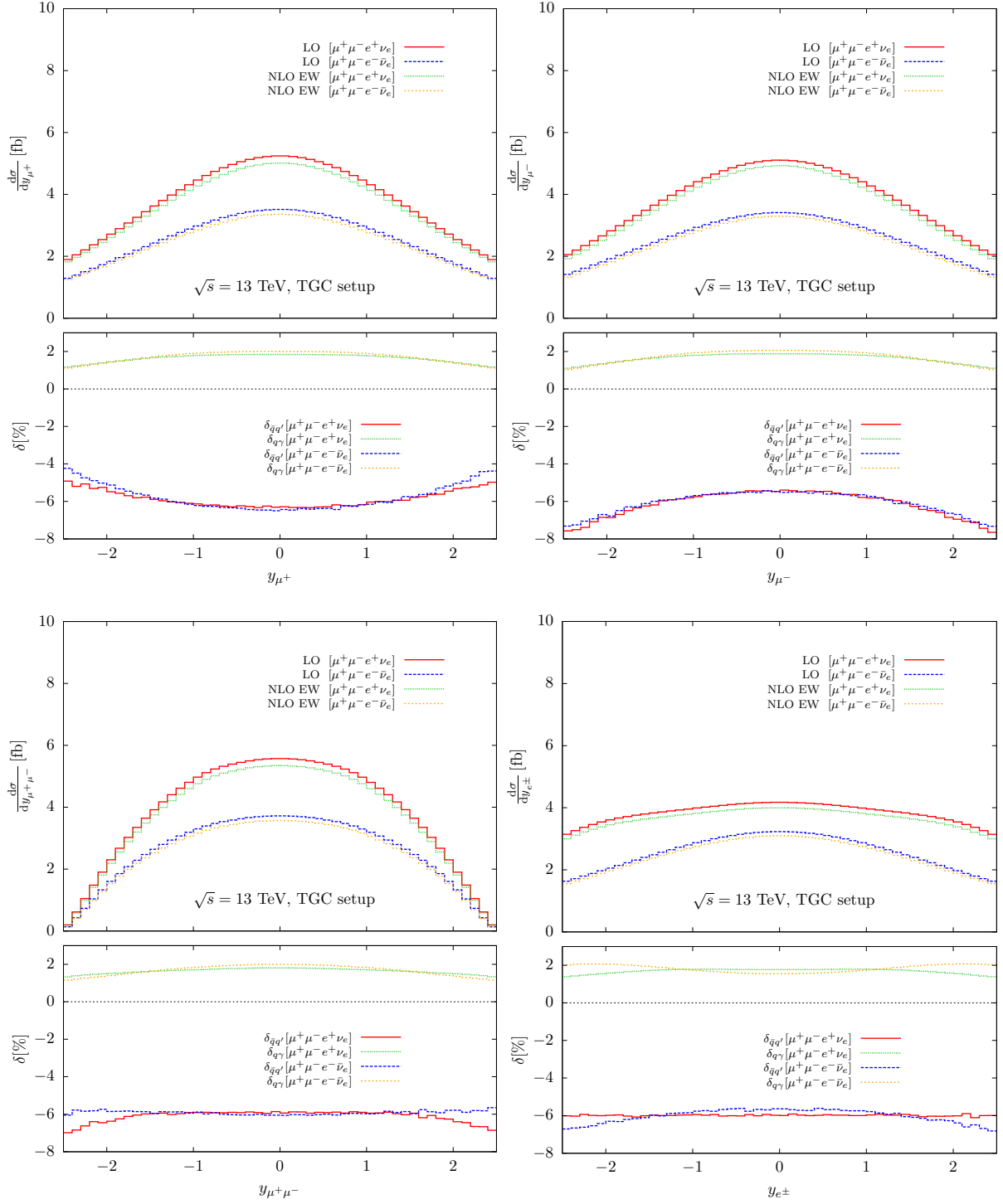
\caption{Distributions in the rapidities of the W and Z decay leptons in the TGC 
setup for $\PZ\PW^+$ and $\PZ\PW^-$: $\mu^+$ (upper left panel), $\mu^-$ (upper right panel), $\mu^+\mu^- $ 
pair (lower left panel) and $\Pe^\pm$ (lower right panel).}
\label{fig:TGC.leptonRapidities}
\efi
Figure~\ref{fig:TGC.leptonRapidities} shows various distributions in
the rapidities of the charged leptons for the mixed-flavour final
states of $\PZ\PW^+$ and $\PZ\PW^-$ in the TGC setup. In the upper
row, distributions in the rapidities of the $\mu^+$ and the $\mu^-$
are presented. The corresponding photon-induced corrections are rather
flat and almost equal for $y_{\mu^+}$ and $y_{\mu^-}$. The
quark-induced corrections show a striking difference in the curvature
of the relative corrections: those for $y_{\mu^+}$ are minimal in the
central region and maximal in forward direction, while it is just the
other way round for $y_{\mu^-}$. In forward direction, the difference
between the corrections to the two observables amounts to $2.5\%$ for
$\PZ\PW^+$ and $3.0\%$ for $\PZ\PW^-$.  This behaviour originates from
the difference in the PDFs of up and down quarks in combination with
the fact that the matrix element is not symmetric under exchange of
the $\mu^+$ and $\mu^-$ momenta.  The lower left plot shows the
distribution in the rapidity of the $\mu^+\mu^-$ pair, \ie the
rapidity of the reconstructed Z~boson. Like in the upper plots, the
photon-induced corrections are flat and almost equal for $\PZ\PW^+$
and $\PZ\PW^-$.
This holds also for the quark-induced corrections in the central
region. In forward direction, we observe a difference between the
corrections to $\PZ\PW^+$ and $\PZ\PW^-$ of about one percent which
can be attributed to the interplay of PDFs and subtracted virtual
corrections.  The lower right plot shows the distribution in the
rapidity of the charged lepton from W~decay. The photon-induced
corrections stay at the level of $2\%$ with sub-percent deviations
between $\PZ\PW^+$ and $\PZ\PW^-$.  Differences of similar size are
also observed in the quark-induced corrections.  Like for
$y_{\mu^+\mu^-}$, we could show that the difference is induced by the
PDFs and largest for the virtual corrections.

\bfi
\begin{center}
\begin{minipage}{0.49\textwidth}
 \includegraphics[width=\textwidth]{plots/{{ATLAStgc.13TeV.rapdiff.samesignwzleptons.2m1e}}}
\end{minipage}
\begin{minipage}{0.49\textwidth}
  \includegraphics[width=\textwidth]{plots/{{ATLAStgc.13TeV.rapdiff.oppsignwzleptons.2m1e}}}
\end{minipage}
\end{center}
\caption{Distribution in the rapidity difference of the $\mu^\pm\Pe^\pm$ pairs (left panel) and the $\mu^\pm\Pe^\mp$ pairs (right panel) for $\PZ\PW^+$ and $\PZ\PW^-$ in the TGC setup.}
\label{fig:TGC.rapdiff.wzleptons}
\efi
Figure~\ref{fig:TGC.rapdiff.wzleptons} displays the distribution in
the rapidity difference of the $\mu^\pm\Pe^\pm$ pairs (left panel) and
in the rapidity difference of the $\mu^\mp\Pe^\pm$ pairs (right panel)
for $\PZ\PW^+$ and $\PZ\PW^-$ in the TGC setup.  The NLO EW
corrections of the quark-induced channels show characteristic
percent-level differences between the $\PZ\PW^+$ and $\PZ\PW^-$ case.
For the same-sign pair $\mu^+\Pe^+$ ($\mu^-\Pe^-$), the corrections in
the $\PZ\PW^+$ case ($\PZ\PW^-$ case) have a maximum at $\Delta y=0$
with $-4\%$ ($-5\%$), and reach $-8\%$ ($-6\%$) at $|\Delta y|=5$.
The maximal difference of the corrections is, thus, largest for the
$\PZ\PW^+$ case. In the opposite-sign case ($\mu^\mp\Pe^\pm$ pairs),
the behaviour is the other way round. The corrections for $\PZ\PW^+$
vary between their extrema by only about $2\%$, while the variation
for $\PZ\PW^-$ amounts to almost $5\%$. The photon-induced corrections
are basically equal for both final states and rather flat.

\section{Conclusions}\label{se:conclusion}

The production of a pair of a neutral and a charged vector boson with
subsequent leptonic decays is a very important process for precision
tests of the Standard Model of elementary particles and in searches
for anomalous triple-gauge-boson couplings (TGC).  In this article,
the first computation of next-to-leading order (NLO) electroweak (EW)
corrections to the production of three charged leptons plus missing
energy at the Large Hadron Collider has been presented. We have
analysed the four independent final states $ \mu^+\mu^- \Pe^+
\nu_\Pe$, $\mu^+\mu^- \Pe^- \bar\nu_\Pe$, $\mu^+\mu^- \mu^+\nu_\mu$,
and $\mu^+\mu^- \mu^- \bar\nu_\mu$ applying realistic experimental
phase-space cuts, first in a rather inclusive setup with minimal event
selection and second in a scenario that is tailored to TGC searches.
We use the complete matrix elements including all off-shell effects of
intermediate massive vector bosons and virtual photons as well as
irreducible background diagrams.

We have computed the NLO EW corrections resulting from
quark--antiquark initial states as well as from initial states with
photons. The photon-induced corrections raise the leading-order (LO)
cross sections by about $+2\%$ in both scenarios. The quark-induced
corrections depend significantly on the fiducial volume.  They lower
the cross section by about $-3\%$ in the inclusive case and by about
$-6\%$ in the TGC setup. For the fiducial cross section, the
corrections to final states with positive total charge ($\PZ\PW^+$)
differ from those with negative total charge ($\PZ\PW^-$) at the
sub-percent level.

At the level of differential distributions, we observe quark-induced
corrections of up to $-30\%$ in the high-energy tails of distributions
stemming from EW Sudakov logarithms. The photon-induced corrections
show an opposite behaviour. They are positive over the whole phase
space and grow with increasing transverse momenta and thus partly
compensate the quark-induced corrections. Since the photon-induced
contributions first occur at NLO for the considered processes, their
impact can efficiently be suppressed by applying an appropriate jet
veto.

Comparing the processes with opposite total charge of the final state
($\PZ\PW^+$ and $\PZ\PW^-$), we observe that the photon-induced
corrections exhibit significant differences of more than $10\%$ for
certain observables. The different behaviour between the $\PZ\PW^+$
and $\PZ\PW^-$ channels results from the differences in the parton
distribution functions for the respective initial states.  Concerning
quark-induced corrections, percent-level differences
between final states with opposite charge are found in rapidity
distributions. Differential distributions that are sensitive to
kinematic thresholds or resonances show typical radiative tails
induced by photon-radiation off final-state leptons and lead to
characteristic differences between the considered inclusive and TGC
setups.

We have studied the impact of interference effects arising in final
states with equal-flavour leptons. In the inclusive setup, they lower
the fiducial cross section with respect to the mixed-flavour case at
the permille level only. While this holds true also at the level of
differential distributions in phase-space regions dominated by on-shell
vector-boson-pair production, interference effects become sizeable and
lower the cross section by up to one third in off-shell
sensitive phase-space regions. If the observables or the phase-space
cuts depend on the selection of equal-flavour final-state leptons
(like in the considered TGC scenario), the differences between equal-
and mixed-flavour processes are, in general, more pronounced and cannot
exclusively be attributed to interferences. Both at LO and NLO, the
shape distortions owing to lepton selection and interferences are
almost equal in size.

The NLO EW corrections for WZ production presented in this article are
important for precision tests of the Standard Model and its possible
extensions. Taking into account that this process is meanwhile known
at next-to-next-to-leading order in the strong coupling, the NLO EW
corrections further reduce the theoretical uncertainty and can help to
improve the exclusion limits on anomalous TGC. We advocate for a
systematic inclusion of the EW corrections in future experimental
analyses.

\subsection*{Acknowledgements}
We would like to thank Stefan Dittmaier for useful discussions. The
work of B.B. and A.D. was supported by the German Federal Ministry for
Education and Research (BMBF) under contract no.~05H15WWCA1 and by the
German Science Foundation (DFG) under reference number DE 623/6-1.
The work of L.H. was supported by the Spanish MINECO grants
FPA2013-46570-C2-1-P and FPA2016-76005-C2-1-P, by the grant
2014-SGR-104, and partially under the project MDM-2014-0369 of ICCUB
(Unidad de Excelencia “Mar\'ia de Maeztu”).

\bibliographystyle{JHEP}                
\bibliography{bibliography}

\end{document}